\documentclass[aps,prl,twocolumn,superscriptaddress,showpacs]{revtex4}
\usepackage{amsmath}
\usepackage{graphicx}

\input{tcilatex}

\begin{document}

\textbf{Comment on ``Elastic Stabilization of a Single-Domain Ferroelectric
State in Nanoscale Capacitors and Tunnel Junctions"}

In a recent Letter \cite{perkohl07}, Pertsev and Kohlstedt (PK) claim that
"even nanoscale capacitors and tunnel junctions may have out of plane
polarization sufficient for memory applications." Here we show in an
elementary way that this conclusion is not substantiated by their
calculations and that they should have come to the opposite conclusion
within their approximations.

Indeed, while obtaining their results for the critical thickness $t_{c,\max
} $ (e.g. Fig.2 \cite{perkohl07})\ Pertsev and Kohlstedt \textquotedblleft
simplified the problem by setting the gradient coefficients $g_{\ln }$ to
zero\textquotedblright . Physically, this means that the domain wall energy
and the width were set to zero. Since PK consider a capacitor with imperfect
metallic electrodes (or nonFE layers near perfect electrodes) there would be
a depolarizing field in a homogeneously polarized film with short-circuited
electrodes, and it costs \emph{nothing} to the system to eliminate\ it by
creating domains when $g_{\ln }=0$. Therefore, they presented results that
refer to states \emph{unstable} with respect to domain formation. Indeed,
the stability of the considered homogeneous states is governed exactly by
those gradient terms that PK have nullified. When interested in a real
situation, one has to use the data or the first-principles calculations to
estimate $g_{\ln }$. An example of such an analysis using neutron scattering
data for BaTiO$_{3}$ can be found in \cite{BLapl06,BLcm06,BLmemory07}. As
follows from the discussion there, a real challenge for the material science
consists in finding electrodes, ferroelectrics, and substrates to ensure a
practical stability of the homogeneous polarization. PK has not even touched
upon this chief problem.

PK believe that the elastic energy associated with the domain structure
prevents its formation. However, this energy definitely plays no role within
their approximation where the equation of state admits solutions with
strictly rectangular distribution of the polarization $P(x)=\pm p,$ $%
P^{2}(x)=p^{2}=\mathrm{const}$. For $p$ equal the polarization of the
homogeneous state, the strains corresponding to this solution are \emph{%
exactly} the same as for the homogeneous state, i.e. the elastic energy
associated with the domains is strictly zero. This does not mean, of course,
that the elasticity plays no role in formation of 180$^{\circ }$ domain
structure, but this role is misunderstood by PK. Hence, the criticism of
prior works, Refs.[17,18] in \cite{perkohl07}, for \textquotedblleft
overlooking\textquotedblright\ the elasticity is unfair: its account is
irrelevant when one defines the point of stability of the paraelectric phase
with respect to the domain formation studied there. Indeed, the stability
loss involves solution of a system of linear equations for the
\textquotedblleft polarization waves\textquotedblright\ while the striction
term provides a nonlinear contribution renormalizing the term $BP^{4}$ in
the thermodynamic potential. This term begins to play a role when the
amplitude of the \textquotedblleft polarization waves\textquotedblright\ has
to be found but the correction introduced are not essential when one
considers a second order transition studied there. The effects of striction
in the paraelectric phase, or, more generally, effects of quadratic coupling
of the order parameter with strains are indeed important when one considers
loss of stability of a nonsymmetric phase with temperature or size-effect
driven phase transition to the symmetrical phase. This was understood long
ago for bulk transitions \cite{Lev74} and should be reconsidered for phase
transitions in strained films, but discussion in Ref.\cite{perkohl07} is
irrelevant there.

The main point of this Comment is not that the calculation in Ref.\cite%
{perkohl07} are incorrect, as they are, but that the PK approach gives no
clue about the possibilities of memory in nanoscale FE capacitors. Indeed,
the question they address (as prior Refs.[17,18]) is a stability with
respect to very small fluctuations. Lack of such stability means, of course,
impossibility of a memory. But the stability can mean two different things:
an absolute stability (absolute minimum of energy) and a relative stability,
i.e. metastability. In the latter case a memory is a question of time of
escape from the metastable state. If this time is too short for
applications, we have the case of practical absence of memory. The authors
do not even mention this possibility. This is surprising given a classic
example in ferroelectricity where the instability of homogeneously polarized
state has little to do with polarization switching. Recall that the field at
which the homogeneously polarized state becomes unstable is one or two
orders of magnitude larger than the experimental coercive field. In other
words, the switching occurs \emph{not} because of instability but because of
domain nucleation and growth.

Similarly, a memory loss in all likelihood is not a question of instability
with respect to infinitely small fluctuations but a question of the domain
nucleation, see \cite{BLmemory07}. We can also refer to the data of Noh's
group \cite{Noh05} where the \textquotedblleft critical thickness for
ferroelectricity\textquotedblright\ and \textquotedblleft the critical
thickness for memory\textquotedblright\ are found to be quite different. The
key theoretical problem to find the \textquotedblleft critical thickness for
FE memory\textquotedblright\ is that of calculating the escape time from
metastable states. One has to look for adequate approaches to the problem of
ferroelectric memory and, unfortunately, Ref. \cite{perkohl07}, with a true
faith bestowed on numerical results without doing a qualitative analysis,
does not help but rather grossly misleads this effort.

A.M. Bratkovsky$^{1}$ and A.P. Levanyuk$^{1,2}$

$^{1}$Hewlett-Packard Labs, Palo Alto, California 94304

$^{2}$Dept. Fiz. Mat. Cond., UAM, Madrid 28049, Spain


\end{document}